\documentclass[twocolumn, tighten, twocolappendix]{aastex631_mod}
\usepackage[T1]{fontenc}
\usepackage{microtype}
\usepackage{newtxtext}
\usepackage[varvw]{newtxmath}
\usepackage{mathtools}
\usepackage{booktabs}
\usepackage{tabularx}
\usepackage{multirow}

\hypersetup{pdfauthor={Errani et al.},
            pdftitle={},
            pdfkeywords={},
            bookmarksnumbered=true}

\newcommand{\rmx}{r_\mathrm{mx}}

\newcommand{\Vmx}{V_\mathrm{mx}}
\newcommand{\Mmx}{M_\mathrm{mx}}

\newcommand{\kms}{\mathrm{km\,s^{-1}}}
\newcommand{\Rh}{R_\mathrm{h}}

\newcommand{\rh}{r_\mathrm{h}}
\newcommand{\Mh}{M_\mathrm{h}}

\newcommand{\diff}{\mathrm{d}}
\newcommand{\Msol}{\mathrm{M_{\odot}}}

\newcommand{\pc}{\mathrm{pc}}
\newcommand{\Gyr}{\mathrm{Gyr}}
\newcommand{\Myr}{\mathrm{Myr}}
\newcommand{\Ydyn}{\Upsilon_\mathrm{dyn}}
\newcommand{\Ydynzero}{\Upsilon_\mathrm{dyn0}}

\newcommand{\sigmalos}{\sigma_\mathrm{los}}

\newcommand{\g}{\mathrm{g}}
\newcommand{\E}{\mathcal{E}}
\newcommand{\plus}[1] {^{\mathmakebox[\widthof{$^-$}][c]{+}#1}}
\newcommand{\minus}[1]{_{\mathmakebox[\widthof{$^-$}][c]{-}#1}}

\graphicspath{{./}{./Figures/}}

\begin{document}
\title{Collisional Dynamics of Stars and Dark Matter in Ultra-Faint Galaxies}
\shorttitle{Collisional Dynamics of Stars and Dark Matter}
\shortauthors{Errani et al.}

\author{Rapha\"el Errani}
\affiliation{McWilliams Center for Cosmology and Astrophysics, Department of Physics, Carnegie Mellon University, Pittsburgh, PA 15213, USA}
\email{errani@cmu.edu}

\author{Nicolas Esser}
\affiliation{Service de Physique Th\'eorique, Universit\'e libre de Bruxelles, Boulevard du Triomphe, CP225, 1050 Brussels, Belgium}

\author{Jorge Pe\~narrubia}
\affiliation{Institute for Astronomy, University of Edinburgh, Royal Observatory, Blackford Hill, Edinburgh EH9 3HJ, UK}

\author{Matthew G. Walker}
\affiliation{McWilliams Center for Cosmology and Astrophysics, Department of Physics, Carnegie Mellon University, Pittsburgh, PA 15213, USA}

\received{April 6 2026}
\revised{June 20 2026}
\accepted{June 22 2026}

\begin{abstract}
We use controlled $N$-body simulations to study the collisional exchange of energy between stars and dark matter in ultra-faint galaxies. We find that dynamical friction between stars and subsolar-mass dark matter particles results in the depletion of dark matter from the galaxies' centers, thereby transforming dark matter cusps into constant-density cores. The process is particularly effective in tidally limited galaxies with low stellar velocity dispersion. As high-mass stars sink toward the center of the dark matter halo, the dynamical-to-stellar mass ratio within the stellar half-light radius decreases monotonically. The stellar population of a dark matter-dominated galaxy is thereby compacted into a dense, baryon-dominated cluster, surrounded by a dark matter halo. Such a cluster would share the chemical composition of an ultra-faint galaxy, yet would be virtually dark matter-free within its half-light radius. We moreover find that the collisional cooling with dark matter particles provides an efficient pathway for the formation of stellar binaries in the contracting cluster. The contraction is eventually slowed down due to the decreasing central dark matter densities and the formation of stellar binaries. Our models highlight that the dynamical processes governing the faintest galaxies give rise to a rich phenomenology, blurring the line between the dynamics of globular clusters and galaxies.
\end{abstract}
\keywords{Cold dark matter (265); Dwarf spheroidal galaxies (420); Dynamical evolution (421); Dynamical friction (422); $N$-body simulations (1083); Star clusters (1567)}

\section{Introduction}
Cold Dark Matter (CDM) cosmology predicts galaxies to form in the deep potential wells of massive dark matter halos \citep{WhiteRees1978}. Dark matter substructure is expected to exist in a hierarchy of halos, subhalos, sub-subhalos, and so forth, down to the smallest clustering scales set by dark matter physics \citep{Tormen1997, Springel2008, Wang2020, Zheng2024}. Cosmological simulations suggest that, in the absence of baryons, cold dark matter halos follow a universal centrally-divergent radial density distribution \citep[][the NFW profile]{Navarro1996a, Navarro1997}. The central high-density ``cusps'' of NFW halos render them highly robust to the effects of galactic tidal forces \citep{Penarrubia2010, EPLG17, vdb2018}: in a smooth tidal field, cuspy subhalos evolve toward a stable asymptotic remnant state and tidal mass loss virtually stalls \citep{EN21, Stucker2023}. Stellar systems embedded in the density cusps of CDM subhalos are therefore protected from full tidal disruption, allowing ultra-faint galaxies to survive even in the strong tidal field of the inner regions of the Milky Way, and thereby plausibly giving rise to a population of heavily-stripped ``micro galaxies'' with luminosities and structural properties at the interface between the globular clusters and dwarf galaxy regimes \citep{EP20, EINPW2024}. In recent years, deep photometric surveys have revealed numerous ``ambiguous'' stellar systems with properties similar to those of the faintest globular clusters and the faintest, smallest galaxies \citep{Conn2018, Cerny22, Cerny23, Mau2020, Simon2024, Smith2024, Tan2026}. A subset of these objects, with old stellar ages, low metallicities, and orbits that reach the very inner regions of the Milky Way, constitute plausible candidates for tidally stripped ``micro galaxies'' \citep{ENSM24, Simon2024, Smith2024, Cerny2026}. 

Traditionally, stars in dwarf galaxies are assumed to obey the collisionless Boltzmann equation, that is, the smooth mean gravitational field of the host galaxy is expected to determine their dynamics. Recent numerical work, however, suggests that this picture is likely incomplete because of the heating effects from dark matter substructures \citep{Penarrubia2025}. Further, for small systems with short crossing times, such as ultra-faint and ``micro galaxies'', the minute potential fluctuations due to the stars' own gravity can accumulate over time and result in the segregation of stellar masses \citep{EPW2025} -- an effect previously thought to be limited to dark matter-free systems and suggested as a litmus test for the absence of dark matter \citep{Kim2015, Baumgardt2022, Simon2024, Zaremba2025}.  

These recent theoretical and observational developments motivate a re-evaluation of the effects of collisional energy exchange between stars and dark matter in the smallest of galaxies. Consider, as a crude estimate of the time scales at play, the Chandrasekhar dynamical friction time scale for point masses in an isothermal potential, adapted from \citet[][equation 7-26]{BT87}
\begin{equation}
 \frac{T_\mathrm{fric}}{\Gyr} ~\sim~  \frac{0.26}{\ln \Lambda} \left(  \frac{r_\mathrm{i}}{\mathrm{pc}} \right)^2 \left( \frac{ v_\mathrm{c}}{\kms}  \right) \left( \frac{m_\star}{\Msol} \right)^{-1}~,
 \label{Eq:FrictTimeScale}
\end{equation}
where by $r_\mathrm{i}$ we denote the initial distance of a star from the galaxy's center, $v_\mathrm{c}$ its circular orbital velocity, and $m_\star$ its mass. Using a Coulomb logarithm\footnote{The choice of $\ln \Lambda=5$ yields results consistent with the $N$-body experiments discussed in Sec.~\ref{Sec:StarClusterFormation}. For comparison, \citet{Fellhauer2007} measure $\ln \Lambda \approx 3.7$ in $N$-body simulations of the orbital decay of a point mass in an isothermal host potential.} of $\ln \Lambda \sim 5$, for classical dwarf galaxies with sizes of order $r_\mathrm{i}\sim100\,\pc$ and typical orbital velocities of order $v_\mathrm{c} \sim10\,\kms$, the above equation suggests a sinking time scale for solar-mass stars that exceeds the age of the universe by a factor of over a hundred. In classical dwarf galaxies, the collisional energy exchange between stars and dark matter has little effect on the dynamical evolution of the stellar component\footnote{By comparison, the orbital decay of globular clusters via dynamical friction in classical dwarf galaxies occurs on time scales of order ${\sim}\Gyr$, and has been suggested as a mechanism to drive globular clusters toward the centers of their hosts \citep[see, e.g.,][]{Hernandez1998, Goerdt2006, Angus2009, Cole2012, Borukhovetskaya2022_Fornax}.}. However, for tidally stripped ultra-faint and ``micro galaxies'' with sizes of $r_\mathrm{i} \sim 10\,\pc$ and typical orbital velocities of order $v_\mathrm{c} \sim 1\,\kms$, we find $T_\mathrm{fric} \sim 5 \,\Gyr$ \citep[see also][Appendix B]{Esser2026} -- substantially lower than the typical age of many of these systems \citep{Cerny2026}, and plausibly crucial for their dynamical evolution. 

In this work, we use controlled $N$-body simulations to study the effects of the collisional energy exchange between stars and dark matter in ultra-faint galaxies. We focus on tidally limited systems, where both the dark matter and the stellar distributions are truncated beyond the luminous radii \citep[see][]{ENIP2022}. With this setup, more than half of the dark matter particles in our simulations fall within the initial stellar half-light radius. This allows us to resolve the dark matter halo with $N$-body particle masses of $m_\mathrm{DM} = 10^{-3}\,\Msol$, and individual stars with stellar masses of $m_\star = 0.2$--$0.8\,\Msol$. As $m_\mathrm{DM} \ll m_\star$, our models directly resolve the dynamical friction exerted by the dark matter on individual stars. We will show that dynamical friction drives high-mass stars toward the center of the ultra-faint galaxy, resulting in the formation of a self-gravitating central stellar cluster. This process injects heat (i.e., random kinetic energy) into the dark matter cusp, which in turn is transformed into a constant-density core. The mechanism underlying the core formation is analogous to that originally proposed by \citealt{El-Zant2001} for dynamical friction between ``lumps'' of gas and dark matter. In the systems considered here, however, the energy exchange is driven by the collisions of subsolar-mass dark matter particles and stars. Crucially, the cusp-to-core transformation driven by dynamical friction between stars and dark matter provides a new pathway for the formation of density cores in systems where the effects of baryons on the dark matter distribution were traditionally expected to be negligible \citep{Penarrubia2012, DiCintio2014, Onorbe2015}.

The paper is structured as follows. In Sec.~\ref{Sec:NumericalSetup}, we discuss the numerical setup of our controlled $N$-body experiments. In Sec.~\ref{Sec:CuspHeating}, we discuss the heating of dark matter cusps by stellar perturbations. We then turn our attention to the dynamical evolution of the stellar component in Sec.~\ref{Sec:StarClusterFormation}, and discuss the formation of stellar binaries in Sec.~\ref{Sec:Binaries}. We study the thermodynamics of the energy exchange between stars and dark matter in Sec.~\ref{Sec:Thermodynamics}. 
Finally, we discuss the observational implications of our results as well as caveats to our analysis in Sec.~\ref{Sec:Discussion}.

\section{Numerical setup}
\label{Sec:NumericalSetup}
With the aim of studying the collisional dynamics between stars and dark matter in ultra-faint galaxies, we perform a series of $N$-body experiments. 

\subsection{Stars}
We model the stellar system to resemble Delve~1, a Milky Way satellite with a total stellar mass of $144\plus{24}\minus{27}\,\Msol$ and a projected (2D) half-light radius of $6.2\plus{1.5}\minus{1.1}\,\pc$ \citep{Mau2020, Simon2024}. As in \citet{EPW2025}, we model the stellar population as a two-component system consisting of high- and low-mass stars, with stellar masses of $0.8\,\Msol$ and $0.2\,\Msol$, respectively. This discretization of the stellar mass function is motivated by the \citet{Chabrier2003} present-day mass function, where half of the stellar mass is in stars below ${\sim}0.5\,\Msol$, with a median mass of ${\sim}0.2\,\Msol$, whereas the remaining half consists of stars with a median mass of ${\sim}0.8\,\Msol$.

Stars are initially distributed according to a spherical exponential profile,
\begin{equation}
\rho_\star(r) = \rho_0  \exp(-r / r_\star)
\label{Eq:ExpProfile}
\end{equation} 
with total mass $ M_\star = 8 \pi \rho_0 r_\star^3 $, where $\rho_0$ denotes the central stellar density and $r_\star$ is a scale radius. This profile has a (3D) half-light radius of $r_\mathrm{h} \approx 2.67\, r_\star$, and a projected (2D) half-light radius of $\Rh \approx 2.02\, r_\star$. The corresponding potential is
\begin{equation}
  \Phi_\star(r)= - G M_\star \left[  1 - (1 + 0.5\, r / r_\star) \exp(-r/r_\star)   \right] / r.
  \label{Eq:StellarPot}
\end{equation}
Both the population of low- and high-mass stars share the same total stellar mass, and the same initial half-light radius. The initial parameters are listed in Table~\ref{Tab:SimParam}. 

We generate equilibrium $N$-body realizations of the two stellar populations in the combined potential of stars (Eq.~\ref{Eq:StellarPot}) and dark matter (Eq.~\ref{Eq:DMPot}), using the Eddington-inversion code \textsc{nbopy} \citep{EP20}, available online\footnote{\url{https://github.com/rerrani/nbopy}}. 

\begin{table}[tb]
\centering
\caption{Simulation parameters. The stellar population is approximated by a two-component system of low-mass and high-mass stars, with a combined stellar mass of $144\,\Msol$. The table lists the initial 3D half-light radius $r_\mathrm{h0} \approx (4/3) R_\mathrm{h0}$, population mass $M_\star$, mass $m_\star$ of a single star, and number of stars $N_\star$. We run simulations for three dark matter halos with initial characteristic size $r_\mathrm{mx0}$, mass $M_\mathrm{mx0}$, particle mass $m_\mathrm{DM}$, and particle number $N_\mathrm{mx0} \equiv N({{<}r_\mathrm{mx0}})$ as listed below, with resulting initial dynamical-to-stellar mass ratios of $\Ydynzero=3$, $10$ and $30$. For reference, we also run a dark matter-only simulation with the same halo parameters as in the $\Ydynzero = 3$ model. Throughout the paper, the initial values listed here are denoted by a subscript zero, e.g., $\Ydynzero \equiv \Ydyn(t=0)$. }
\begin{tabularx}{\linewidth}{l@{\hspace{0.85cm}}c@{\hspace{0.4cm}}c@{\hspace{0.4cm}}c@{\hspace{0.6cm}}c@{\hspace{0.5cm}}c}
 \toprule
 \textsc{Stars}           & $r_\mathrm{h0}/\mathrm{pc}$   & $M_\star/\mathrm{M}_\odot$        & $m_\star/\mathrm{M}_\odot$           & $N_\star$             & profile \\ \midrule
 low-mass                 & \multirow{2}{*}{8.3}          & \multirow{2}{*}{72}               & $0.2$                                & $360$                 & \multirow{2}{*}{Eq.~\ref{Eq:ExpProfile}}   \\
 high-mass                &                               &                                   & $0.8$                                & $90$                  & \\ \midrule
\end{tabularx}
\begin{tabularx}{\linewidth}{l@{\hspace{0.2cm}}c@{\hspace{0.15cm}}c@{\hspace{0.15cm}}c@{\hspace{0.2cm}}c@{\hspace{0.2cm}}c}
 \textsc{Dark Matter}     & $r_\mathrm{mx0}/\mathrm{pc}$  & $M_\mathrm{mx0}/\mathrm{M}_\odot$ & $m_\mathrm{DM}/\mathrm{M}_\odot$     & $N_\mathrm{mx0}$      & profile\\ \midrule
 $\Ydynzero = 3$          & \multirow{3}{*}{8.3}          & 144                               & \multirow{2}{*}{$1\!\times\!10^{-3}$}& $1.4 \!\times\! 10^5$ &  \multirow{3}{*}{Eq.~\ref{Eq:ExpCuspProfile}}      \\
 $\Ydynzero = 10$         &                               & 648                               &                                      & $6.5 \!\times\! 10^5$ &     \\
 $\Ydynzero = 30$         &                               & 2088                              &  $4\!\times\!10^{-3}$                & $5.2 \!\times\! 10^5$ &     \\
 \bottomrule
\end{tabularx}
\label{Tab:SimParam}
\end{table}

\subsection{Dark Matter}
We embed the two stellar components in a dark matter subhalo, modelling the example system Delve~1 as a tidally-limited dwarf, motivated by its old stellar population and present-day location in the inner region of the Milky Way \citep{Simon2024}. For tidally limited dwarfs, the radial extent of the surrounding dark matter subhalo is sharply truncated beyond the stellar half-light radius ($\rmx \approx \rh$, see \citealt{ENIP2022} for details). The radial dark matter density profile is well-approximated by an exponentially truncated NFW cusp \citep{EN21}, 
\begin{equation}
 \rho_\mathrm{DM}(r) = \rho_\mathrm{s} r_\mathrm{s} \exp(-r / r_\mathrm{s}) / r~,
\label{Eq:ExpCuspProfile}
\end{equation}
with a total mass $ M_\mathrm{DM} = 4 \pi \rho_\mathrm{s} r_\mathrm{s}^3 $, where $\rho_\mathrm{s}$ and $r_\mathrm{s}$ denote a scale density and scale radius, respectively.
The corresponding circular velocity profile peaks at a radius $\rmx = 1.79\,r_\mathrm{s}$ with a peak squared velocity $\Vmx^2 = 0.30\,G M_\mathrm{DM}/r_\mathrm{s}$, and the mass enclosed within $\rmx$ equals $\Mmx = 0.54 \, M_\mathrm{DM}$. The dark matter density sources the potential:
\begin{equation}
 \Phi_\mathrm{DM}(r)= - G M_\mathrm{DM} \left[1 - \exp(-r/r_\mathrm{s}) \right] / r ~.
 \label{Eq:DMPot}
\end{equation}
For later reference, we also provide the radial (1D) velocity dispersion profile, computed under the assumption of isotropic kinematics (and in the absence of baryons):
\begin{equation}
\begin{split}
\sigma^2_{r, \mathrm{DM}}(r) = G M_\mathrm{DM} \, [  ({r}/{r_\mathrm{s}})~ \mathrm{exp}({r}/{r_\mathrm{s}})~E_1( {r}/{r_\mathrm{s}})  \\ + ({r_\mathrm{s}}/{r}) - ({r_\mathrm{s}}/{r})~  \mathrm{exp}(-{r}/{r_\mathrm{s}}) -1 ]  / (2\,r_\mathrm{s})~,
\end{split}
\label{Eq:ExpCuspSigma}
\end{equation}
where $E_1(r/r_\mathrm{s})$ denotes the exponential integral.

We generate equilibrium $N$-body realizations following the density profile of Eq.~\ref{Eq:ExpCuspProfile} in the combined potential of dark matter and stars. As the dark matter profile is sharply truncated beyond the stellar half-light radius, more than half of the dark matter particles in the models initially lie within the half-light radius (see $N_\mathrm{mx}$ listed in Table~\ref{Tab:SimParam}). This allows us to generate $N$-body models with dark matter particle masses of only $m_\mathrm{DM} = 10^{-3} \, \Msol$, while maintaining modest total particle numbers. The resulting ratio of dark-matter-to-stellar particle mass of $m_\mathrm{DM} / m_\star = 1/200$ (for the case of low-mass stars) and $1/800$ (for the high-mass stars) enables us to resolve the effects of dynamical friction between dark matter and stars without relying on sub-grid prescriptions\footnote{For the $\Ydynzero=30$ model listed in Table~\ref{Tab:SimParam}, we use a slightly larger dark matter particle mass of $m_\mathrm{DM} = 4\times10^{-3}\, \Msol$, which results in ratios of dark-matter-to-stellar particle mass of $m_\mathrm{DM} / m_\star = 1/50$ (low-mass stars) and $1/200$ (high-mass stars).}. 

In this study, we run simulations for three different initial dark matter halo masses, scaled so that the average dynamical-to-stellar mass ratio within the half-light radius $\rh$,
\begin{equation}
 \Ydyn \equiv  1 + M_\mathrm{DM}({<}\rh) / M_\star({<}\rh) ~,
\end{equation}
equals $3$, $10$, and $30$, respectively. The resulting dark matter halo masses and particle numbers are listed in Table~\ref{Tab:SimParam}.

For $\Ydyn=3$, $10$, and $30$, our models of Delve~1 have initial line-of-sight (1D) velocity dispersions of $\langle \sigmalos^2 \rangle^{1/2} \approx (G M_\star \Ydyn r_\mathrm{h}^{-1}/6)^{1/2} \approx 0.2\,\kms$, $0.4\,\kms$ and $0.6\,\kms$, respectively. These values are below the current observational upper limits on the line-of-sight velocity dispersion listed in \citet{Simon2024}, who find that $\langle \sigmalos^2 \rangle^{1/2} \lesssim 1.2\,\kms$ when using a logarithmic prior, and $\langle \sigmalos^2 \rangle^{1/2} \lesssim 2.5\,\kms$ when using a uniform prior. 

\begin{figure*}[t]
 \centering
 \includegraphics[width=\textwidth]{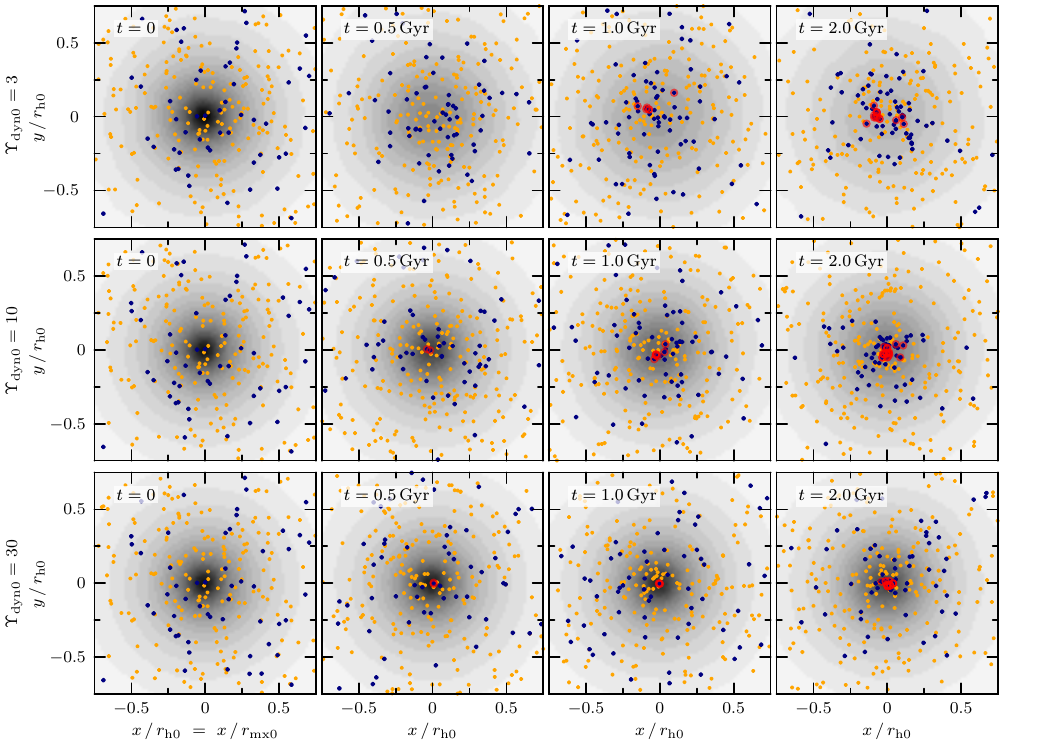}
\caption{Collisional energy exchange between stars and dark matter transforms a dark matter cusp into a constant-density core, as high-mass stars sink to the center of the ultra-faint galaxy. Top row: Simulation snapshots taken at $t=0$, $0.5$, $1$ and $2\,\Gyr$ with parameters as listed in Table~\ref{Tab:SimParam}, for an initial dynamical-to-stellar mass ratio of $\Ydynzero = 3$. The dark matter surface density is shown in gray (darker shades indicating higher density), while high- ($m_\star=0.8\,\Msol$) and low-mass stars ($m_\star = 0.2\,\Msol$) are shown as blue and yellow points, respectively. As high-mass stars sink toward the galaxy's center, the population of low-mass stars approximately retains its original extension, resulting in a segregation of stellar masses within the dwarf galaxy. Central and bottom panels: same as top panel, but for $\Ydynzero = 10$ and $30$, respectively. With increasing $\Ydynzero$, the cusp-to-core transformation becomes less efficient. Binaries with semi-major axis $a_\mathrm{bin} \leq 1\,\pc$ are circled in red. A video version of this figure for the $\Ydynzero=10$ model is available as an \href{https://arxiv.org/src/2604.06304v2/anc}{arXiv ancillary file} (see footnote~\ref{footnote:ProjectionVideoDetails} for details). }
\label{Fig:projected_densities}
\end{figure*} 

\subsection{$N$-body code}
\label{Sec:NbodyCode}
We use the $N$-body code \textsc{gadget-4} \citep{Springel2021_Gadget4} to model the time evolution of stars and dark matter in their combined potential. The force calculation is done using the classical \citet{Barnes1986} oct-tree implemented in \textsc{gadget-4}, adopting a constant (Plummer-equivalent) force softening of $\epsilon_\mathrm{DM}=0.1\,\pc$ for dark matter particles, and $\epsilon_\star=0.02\,\pc$ for interactions between stars. The softened potential is exactly Newtonian for radii larger than 2.8 times the softening length (see equation~71 in \citealt{Springel_2001_Gadget1}). Gravity between dark matter and stars is softened with $\epsilon_\mathrm{DM}$.
To limit spurious heating by dark matter particles, the softening length $\epsilon_\mathrm{DM}$ is chosen to roughly match the mean inter-particle distance $\rmx ( N_\mathrm{mx} )^{-1/3}  \sim 0.1\,\pc $.
To test for convergence, we have degraded the softening lengths by a factor of 2 without any qualitative impact in our results. 
We have chosen conservative values for the dimensionless parameters that determine the force and time integration accuracy of $\alpha = 10^{-3}$ (\texttt{ErrTolForceAcc}) and $\eta = 10^{-3}$ (\texttt{ErrTolIntAccuracy}). The first of these parameters controls the relative cell-opening criterion (through equation~12 of \citealt{Springel2021_Gadget4}). The second parameter sets the time step for a particle $\Delta t = \mathrm{min}[0.5\,\Myr,  (2 \eta \epsilon/|a|)^{1/2}]$, where $\epsilon$ denotes the softening length, and $|a|$ the magnitude of the acceleration (see equation 34~in \citealt{Springel2005Gadget}). 

Note that our simulations are not designed to accurately track the details of the dynamical evolution of the self-gravitating star cluster which forms through the effects of dynamical friction on high-mass stars in our models (see Sec.~\ref{Sec:StarClusterFormation}). The code adopts a softening length for gravitational interactions, and consequently we cannot reliably resolve collisional interactions at scales smaller than the applied softening, such as the formation of close binaries. We therefore end our simulations once the cluster of massive stars becomes fully self-gravitating, i.e., once the dynamical-to-stellar mass ratio within the cluster reaches $\Ydyn\sim1$. For our models with initial dynamical-to-stellar mass ratios $\Ydynzero =3$, $10$, and $30$, we end the simulations after $3\,\Gyr$, $4\,\Gyr$, and $7\,\Gyr$ of evolution, respectively. We ensure that the total energy of the combined system of stars and dark matter drifts by less than $1\,$per cent with respect to the initial value over the simulated period. 

To reduce the impact of discreteness noise, for each choice of $\Ydynzero$, as well as for the dark matter-only model, we run simulations of four random $N$-body realizations of the initial particle distribution. We save simulation snapshots every $10\,\Myr$ of simulated time.

\footnotetext{The video supplementing Figure~\ref{Fig:projected_densities} shows the evolution of stars and dark matter in the $\Ydynzero=10$ model for $2.5\,\Gyr$. File size $13.8\,\mathrm{MiB}$, dimensions $1920\times1920\,\mathrm{px}^2$, duration $28\,\mathrm{s}$. For the sake of brevity, the playback speed is increased between seconds $9$ ($0.1\,\Gyr$) and $19$ ($2.4\,\Gyr$).\label{footnote:ProjectionVideoDetails}}

\section{Results}

\begin{figure*}[t]
 \centering
 \includegraphics[width=\columnwidth]{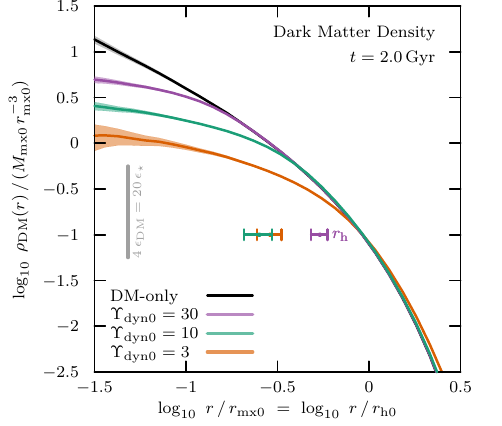} \hfill  \includegraphics[width=\columnwidth]{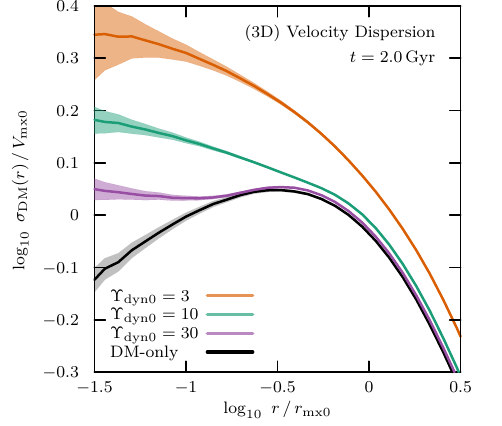} 
 \caption{The initial dark matter cusp is transformed into a constant-density core. Left panel: Radial dark matter density profile $\rho_\mathrm{DM}(r)$ for the simulations with an initial dynamical-to-stellar mass ratio of $\Ydynzero=3$ (orange), $10$ (green) and $30$ (purple). For comparison, the dark matter-only run is shown in black. The smaller $\Ydynzero$ is, the larger the size of the constant-density core. To reduce noise, for each value of $\Ydynzero$, we have stacked the snapshots of all four (random) $N$-body realizations in the interval $t=2\pm 0.1\,\Gyr$. Shaded bands show the Poisson noise level of the stacked snapshots, amplified by a factor of $10$ for visibility. For reference, the half-light radii $r_\mathrm{h}$ of the high-mass stars are indicated using horizontal error bars, and a vertical gray line corresponds to a radius of $4\,\times$ the gravitational softening length for dark matter particles $\epsilon_\mathrm{DM}=0.1\,\pc$. Right panel: Dark matter (3D) velocity dispersion $\sigma_\mathrm{DM}(r)$. As heat (i.e., random kinetic energy) is transferred from the stars to the dark matter, the dark matter velocity dispersion in the central regions increases. Shaded bands show $4\times$ the Poisson noise level of the stacked snapshots. A video supplementing this figure for the $\Ydynzero=10$ model is available as an \href{https://arxiv.org/src/2604.06304v2/anc}{arXiv ancillary file} (see footnote~\ref{footnote:DensityVideoDetails} for details). }
\label{Fig:core_density}
\end{figure*} 

\subsection{Heating of the dark matter cusp}
\label{Sec:CuspHeating}
In the following, we will discuss how collisions between subsolar-mass dark matter particles and stars transfer heat (i.e., random kinetic energy) to the dark matter cusp, which in turn flattens. The mechanism is analogous to the one originally proposed in \citealt{El-Zant2001} for the case of dynamical friction between ``lumps'' of gas and dark matter. Figure~\ref{Fig:projected_densities} shows simulation snapshots at $t=0$, $0.5$, $1$ and $2\,\Gyr$ (left to right) for the three models with initial dynamical-to-stellar mass ratios of $\Ydynzero=3$, $10$ and $30$ (top to bottom). The dark matter surface density is color-coded in gray, with darker shades corresponding to higher surface densities. To reduce Poisson noise when computing the dark matter surface densities, for each choice of $\Ydynzero$, we have stacked the corresponding snapshots of all four random $N$-body realizations (see Sec.~\ref{Sec:NumericalSetup}). The locations of high-mass stars ($m_\star = 0.8\,\Msol$) and low-mass stars ($m_\star = 0.2\,\Msol$) are shown using blue and yellow points, respectively, and are taken from a single $N$-body realization. For the model with $\Ydynzero=3$ (top row), as time progresses, dark matter is depleted from the central regions as high-mass stars sink toward the center due to dynamical friction. The depletion of dark matter progresses more slowly for the models with initial dynamical-to-stellar mass ratios of $10$ and $30$ (central and bottom row). A video version of Fig.~\ref{Fig:projected_densities} for the model with $\Ydynzero=10$ is available as an arXiv ancillary file.

\footnotetext{The video supplementing Figure~\ref{Fig:core_density} shows the time evolution of the enclosed mean dark matter density profile $\bar \rho_\mathrm{DM}({<}r)$ in the $\Ydynzero=10$ model for $2.5\,\Gyr$. To guide the eye, the video also shows the mean stellar densities $\bar \rho_\star({<}\rh)$ of the populations of high-mass and low-mass stars enclosed within their respective half-light radii $\rh$. File size $1.6\,\mathrm{MiB}$, dimensions $1920\times1920\,\mathrm{px}^2$, duration $16\,\mathrm{s}$.\label{footnote:DensityVideoDetails}}

\emph{Cusp-to-core transformation.} 
To quantify the depletion of dark matter seen in Fig.~\ref{Fig:projected_densities}, in the left-hand panel of Fig.~\ref{Fig:core_density}, we show the radial dark matter density profiles $\rho_\mathrm{DM}(r)$ for the $t=2\,\Gyr$ snapshot (beyond $2\,\Gyr$, the dark matter densities evolve only weakly). For reference, the density profile in a dark matter-only control run (with a dark matter particle number matching that of the $\Ydynzero=3$ model) is shown in black. Density profiles corresponding to the $\Ydynzero=3$, $10$ and $30$ models are shown in orange, green and purple, respectively. The depletion of the central density is smallest for the most dark matter-dominated model ($\Ydynzero=30$), and largest for the model with $\Ydynzero=3$. We estimate the size of the constant-density core by fitting an empirical density profile to the $N$-body data. To reduce Poisson noise, we stack $N$-body snapshots from the time interval $t=(2\pm0.1)\,\Gyr$ of the four random $N$-body realizations (see Sec.~\ref{Sec:NumericalSetup}). We use a fitting function of the form 
\begin{equation}
 \rho_\mathrm{core}(r) = \rho_\mathrm{s} r_\mathrm{s} \exp(-r/r_\mathrm{s}) ( r^2 + r_\mathrm{c}^2 )^{-1/2}~,
 \label{Eq:Core}
\end{equation}
where $r_\mathrm{c}$ determines the size of the constant-density core, and $r_\mathrm{s}$, $\rho_\mathrm{s}$ are a scale density and scale radius, respectively. For $r_\mathrm{c}=0$, Eq.~\ref{Eq:Core} reduces to the exponentially truncated cusp of Eq.~\ref{Eq:ExpCuspProfile}. To facilitate the comparison of core sizes $r_\mathrm{c}$, when fitting, we fix $r_\mathrm{s}$ in the fitting function to coincide with the initial scale radius $r_\mathrm{s0} \approx r_\mathrm{h0} / 1.79$ of the exponentially truncated cusp. With this approach, the best-fitting values for the core size are $r_\mathrm{c}/r_\mathrm{s} \approx 0.2$, $0.4$ and $1.0$ for the models with $\Ydynzero = 30$, $10$ and $3$, respectively. 

\emph{Dark matter velocity dispersion.}
As the dark matter cusp is transformed into a constant-density core and the high-mass stars sink toward the center of the dark matter halo, the central velocity dispersion of the dark matter rises. This is shown in the right-hand panel of Fig.~\ref{Fig:core_density}. Note that in the dark matter-only run, the velocity dispersion decreases toward the center of the subhalo, as is typical of cuspy models with isotropic kinematics. For our simulations that include stars, the central dark matter velocity dispersion increases as heat (i.e., kinetic energy) is transferred from the stars to the dark matter. As before, the most dark matter-dominated model ($\Ydyn=30$) sees the smallest increase in central velocity dispersion, whereas the model with $\Ydyn=3$ sees the largest increase. Throughout the heating of the dark matter cusp, the dark matter halo remains approximately in virial equilibrium ($2K+W\approx0$, where $K$ denotes the kinetic and $W$ the potential energy), and maintains isotropic kinematics in the center. The outskirts of the halo, however, develop radially biased kinematics. This is consistent with the picture that dark matter particles are ejected from the galaxy's center as the central dark matter densities decrease. 

\begin{figure}[t]
 \includegraphics[width=\columnwidth]{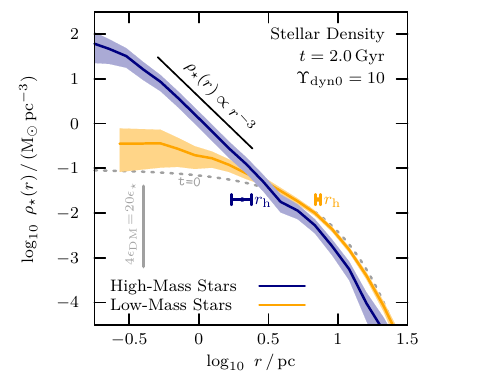}
 \caption{As the population of high-mass stars contracts within the dark matter halo, the central stellar density rises by three orders of magnitude, and the stellar density profile steepens. Shown is the radial stellar density $\rho_\star(r)$ for high-mass stars (blue) and low-mass stars (yellow) at $t=2\,\Gyr$ for the model with $\Ydynzero=10$. To reduce Poisson noise, we have stacked snapshots of all four $N$-body realizations in the interval $(2\pm0.1)\,\Gyr$. In the inner regions, the contracted density profile has a logarithmic slope of $\diff \ln \rho_\star / \diff \ln r \approx -3$. By comparison, the density profile of the low-mass stars has evolved only weakly with respect to the initial configuration (Eq.~\ref{Eq:ExpProfile}, dotted gray curve). For reference, we show the evolved half-light radii $\rh$ of the two populations using horizontal error bars.}
 \label{Fig:StellarDensities}
\end{figure}

\subsection{Formation of a self-gravitating star cluster}
\label{Sec:StarClusterFormation}
We will now turn our attention to describing the dynamical evolution of the stellar component as it contracts within the dark matter halo. As shown in Fig.~\ref{Fig:projected_densities}, the population of high-mass stars sinks toward the center of the dark matter subhalo due to dynamical friction with dark matter particles. 

\emph{Stellar densities.} 
The central stellar densities of the population of high-mass stars increase substantially as the half-light radius contracts. Taking the $\Ydynzero=10$ model as an example, Fig.~\ref{Fig:StellarDensities} shows the radial stellar density profile $\rho_\star(r)$ at $t=2\,\Gyr$ for the high-mass stars (blue) and the low-mass stars (yellow). The density profile of the high-mass stars steepens toward the center, with a logarithmic slope of $\diff \ln \rho_\star / \diff \ln r \approx -3$ (slightly steeper than the slope predicted for dark matter-free, one-component stellar clusters that have undergone core-collapse, see, e.g., \citealt{Cohn1980}, \citealt{GierszHeggie1994}). In contrast, the density profile of the population of low-mass stars has evolved only slightly compared to the initial configuration (Eq.~\ref{Eq:ExpProfile}, shown for reference as a dotted gray curve).

\begin{figure}[t]
 \includegraphics[width=\columnwidth]{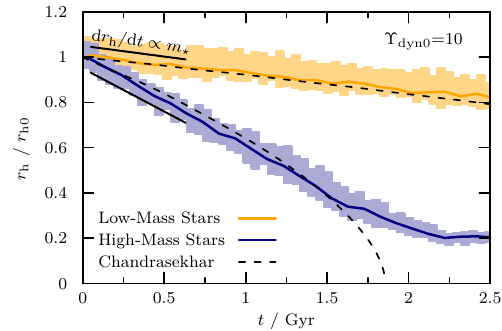}
 \caption{Dynamical friction exerted by dark matter particles on stars drives the contraction of stellar half-light radii. Shown is the evolution of the half-light radii $\rh$ for the two stellar populations of low-mass and high-mass stars (yellow and blue lines, showing the median over all four random $N$-body realizations; shaded bands span the $16^\mathrm{th}$--$84^\mathrm{th}$ percentile of the underlying distribution) for the $\Ydynzero=10$ model. High-mass stars sink to the center more rapidly than low-mass stars. During the early stages of the evolution, the rate of contraction is roughly proportional to the mass of individual stars, $\diff \rh / \diff t \propto m_\star$, consistent with classical Chandrasekhar dynamical friction. For reference, a dashed black curve shows the orbital decay of a star on a circular orbit in a static cuspy potential, computed using the Chandrasekhar dynamical friction formula (see text for details). }
 \label{Fig:DynFrictTime}
\end{figure}

\emph{Dynamical friction time scale.}
Figure~\ref{Fig:DynFrictTime} shows the evolution of the half-light radii $\rh$ of the stellar populations in the model with $\Ydynzero=10$, normalized to their initial values $r_\mathrm{h0}$. Over a period of $2\,\Gyr$, the half-light radius of the population of high-mass stars (shown in blue) decreases by ${\sim}80$ per cent. During the same time, the half-light radius of the low-mass stars (yellow) decreases\footnote{The evolution of the $\Ydynzero=30$ model is qualitatively similar to the $\Ydynzero=10$ case. In contrast, for the $\Ydynzero=3$ model, the half-light radius of the population of low-mass stars increases as the population of high-mass stars contracts: the collisional interaction between the two stellar populations outweighs the effects of dynamical friction between dark matter and low-mass stars.} by only ${\sim}20$ per cent. In the early stages of the evolution, the rate of decay of the half-light radii is roughly proportional to the mass of individual stars, $\diff \rh / \diff t \propto m_\star$. 

For a simple analytical estimate, we use the classical Chandrasekhar dynamical friction formula \citep{Chandrasekhar1941, Chandrasekhar43} to compute the orbital decay of a star in the (cuspy) dark matter density distribution given by Eq.~\ref{Eq:ExpCuspProfile}. Chandrasekhar friction predicts an acceleration opposing the motion (see, e.g., \citealt{BT87} equation 7-18)
\begin{equation}
 \frac{\diff v_\star}{\diff t} = - \frac{4 \pi \ln \Lambda G^2 \rho_\mathrm{DM} m_\star}{v_\star^2} \left[  \mathrm{erf}(\mathcal{X}) - \frac{2\mathcal{X}  \exp(-\mathcal{X}^2)  }{\sqrt{\pi}}  \right]
\label{Eq:FrictionAcc}
\end{equation}
where $\mathcal{X} \equiv v_\star / (\sqrt{2}\, \sigma_{r,\mathrm{DM}})$, and $v_\star$, $m_\star$ denote the orbital velocity and mass of the star. In the above equation, the dark matter density (Eq.~\ref{Eq:ExpCuspProfile}) and velocity dispersion (Eq.~\ref{Eq:ExpCuspSigma}) are denoted by $\rho_\mathrm{DM}$ and $\sigma_{r,\mathrm{DM}}$, respectively. For the sake of simplicity, we manually choose a constant Coulomb logarithm of $\ln \Lambda = 5$. We numerically integrate the orbit of a star on a circular orbit with initial orbital radius equal to $r_\mathrm{h}$ in the dark matter potential of Eq.~\ref{Eq:DMPot}, subject to the friction acceleration of Eq.~\ref{Eq:FrictionAcc}. The results of this rough model are shown as dashed curves in Fig.~\ref{Fig:DynFrictTime}. At early times, the model accurately describes the orbital decay seen in the $N$-body simulations. However, at later times, the model overestimates the rate of orbital decay, in particular for the case of the high-mass stars. The reasons for this discrepancy are twofold. First, the model assumes a static cuspy dark matter density distribution, whereas a constant-density dark matter core forms in the simulation on a time scale of ${\sim}1\,\Gyr$. This dark matter core has a reduced central density compared to the original profile, and a higher central velocity dispersion (see Fig.~\ref{Fig:core_density}) -- both factors that decrease the efficiency of dynamical friction\footnote{Note also that the classical Chandrasekhar friction formula is known to over-estimate the rate of orbital decay in cored dark matter density distributions (see, e.g., \citealt{Read2006}, \citealt{Banik2022}).}. Second, as we will detail in Sec.~\ref{Sec:Binaries}, during the later stages of the evolution, the formation of binary stars in the cluster further slows down its contraction. Despite these complications, classical Chandrasekhar friction between single stars and dark matter provides a surprisingly accurate description of the contraction of the cluster for the first ${\sim}\Gyr$ of evolution.

\begin{figure}[t]
 \centering
 \includegraphics[width=\columnwidth]{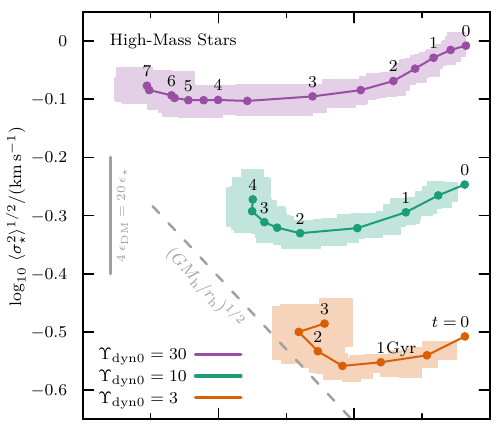}
 
 \includegraphics[width=\columnwidth]{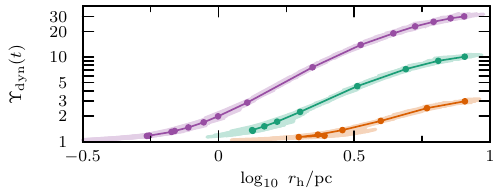} 
\caption{Evolution of the half-light radius $\rh$ and of the average (3D) velocity dispersion $\langle \sigma^2_\star \rangle^{1/2} = \sqrt{3} \langle \sigma^2_\mathrm{los} \rangle^{1/2} $ for the population of high-mass stars (top panel). Models with initial dynamical-to-stellar mass ratios of $\Ydynzero=3$, $10$, and $30$ are shown in orange, green and purple, respectively. The stellar population cools down and contracts, as heat (i.e., kinetic energy) is transferred from stars to dark matter particles. The dynamical-to-stellar mass ratio $\Ydyn$ decreases in this process (see bottom panel). As $\Ydyn$ approaches unity, the stars become self-gravitating, and in turn, the stellar population heats up as it contracts further.}
\label{Fig:HighMassStarsRV}
\end{figure} 

\emph{Evolution of size and velocity dispersion.}
As the population of high-mass stars contracts within the dark matter-dominated potential, it initially cools down. This is the expected behavior for a stellar population that is gravitationally sub-dominant in a dark matter halo \citep[see, e.g.,][]{EPW2025, Penarrubia2025}. Figure~\ref{Fig:HighMassStarsRV} shows the evolution of the half-light radius $\rh$ and the average (3D) velocity dispersion $\langle \sigma_\star^2 \rangle^{1/2} =  \sqrt{3} \langle \sigmalos^2 \rangle^{1/2}  $ for the high-mass stars in the three simulations with initial $\Ydynzero=3$, 10 and 30. As the half-light radius of the high-mass stars contracts, the dynamical-to-stellar mass ratio $\Ydyn(t)$ decreases: the stellar mass becomes increasingly more important for the stellar dynamics (bottom panel of Fig.~\ref{Fig:HighMassStarsRV}). Once the stars become gravitationally dominant, the stellar velocity dispersion grows, and the contraction of $\rh$ progresses at a slower pace until it eventually stalls. 

To guide the eye, a dashed gray line shows the relation $\langle \sigma_\star^2 \rangle^{1/2} \approx (G M_\mathrm{h}/\rh)^{1/2}$, where $M_\mathrm{h} = M_\star/2$ denotes the stellar mass enclosed within the half-light radius $\rh$. This relation approximates the virial\footnote{For a star cluster devoid of dark matter with the density profile of Eq.~\ref{Eq:ExpProfile}, the virial theorem \citep[see, e.g.][]{Amorisco2012, EPW18} predicts a velocity dispersion of $\langle \sigma_\star^2 \rangle = 3 \langle \sigma^2_\mathrm{los} \rangle = (15/96)\,G M_\star/r_\star \approx 0.83\, G \Mh / \rh $.} velocity dispersion expected for a self-gravitating star cluster with an exponential density profile in equilibrium in the absence of binaries. The same relation appears in commonly adopted mass estimators for dwarf spheroidal galaxies, e.g., in the 3D version of the \citet{Wolf2010} mass estimator. We note that even in the regime where the cluster is self-gravitating, $\Ydyn(t) {\sim} 1$, the velocity dispersion of the cluster is substantially higher than predicted by the simple virial relation. In part this can be attributed to the evolving virial coefficient as the star cluster density profile changes shape \citep{Splawska2026}. Crucially, as we will show in Sec.~\ref{Sec:Binaries}, stellar binaries form in the contracting cluster, which add to the observed velocity dispersion\footnote{For reference, the circular orbital velocity (with respect to the center of mass) of an equal-mass binary consisting of two $m_\star=0.8\,\Msol$ stars, with separation equal to the (stellar) force-softening length $\epsilon_\star=0.02\,\pc$, equals $v_\mathrm{bin} = \sqrt{  G m_\star / (2 \epsilon_\star)  } \approx 0.3 \,\kms$.}.  

The aim of the present work is to study the contraction of the stellar population due to dynamical friction. The long-term evolution of the self-gravitating cluster of massive stars lies beyond the scope of the present work (and requires a different numerical setup to overcome the limitations discussed in Sec.~\ref{Sec:NbodyCode}). We therefore end our simulations once $\Ydyn(t) {\sim} 1$, and leave the subsequent evolution to future study. 

\begin{figure}[t]
 \centering
 \includegraphics[width=\columnwidth]{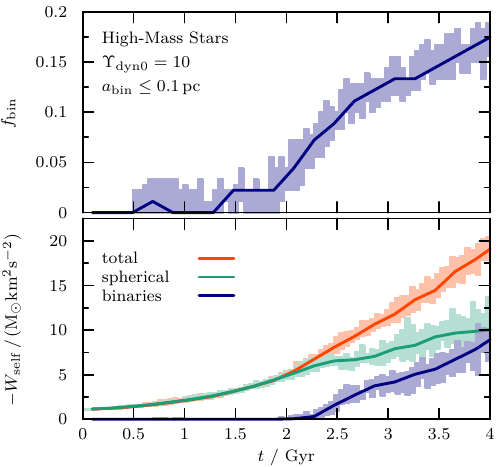}
\caption{As the population of high-mass stars contracts, the number of dynamically formed binaries $N_\mathrm{bin}$ increases. Top panel: Shown is the binary fraction $f_\mathrm{bin} \equiv N_\mathrm{bin} / N_\star$ for binaries consisting of two high-mass stars as a function of time. For the model with $\Ydynzero=10$, the number of binaries with semi-major axes $a_\mathrm{bin} \leq 0.1\,\pc$ rises sharply around ${\sim}2\,\Gyr$. Shaded bands span the $16^\mathrm{th}$--$84^\mathrm{th}$ percentile of the underlying distribution. Bottom panel: Potential energy $W_\mathrm{self} = \sum m_\star \Phi_\star/2$ for the system of high-mass stars as a function of time. Shown is the total potential computed through direct summation (orange curve) and under the assumption of spherical symmetry (green curve). After ${\sim}2\,\Gyr$ of evolution, as the binary fraction $f_\mathrm{bin}$ grows, the contribution by binary stars to $W_\mathrm{self}$ increases (blue curve). }
\label{Fig:Binaries}
\end{figure}  

\subsection{Dynamical formation of stellar binaries}
\label{Sec:Binaries}
In the following, we discuss the formation of stellar binaries in the contracting population of massive stars. As the population of massive stars contracts, the mean stellar density increases: For the model with $\Ydynzero=10$, over $2\,\Gyr$ of evolution, the mean stellar density $\langle \rho_\star ({<}\rh) \rangle$ enclosed within the half-light radius of the population of massive stars rises from its initial value of ${\sim}3 \times 10^{-2}\,\Msol\,\pc^{-3}$ to a value of $1.5\,\Msol\,\pc^{-3}$. At the same time, the stellar velocity dispersion decreases. In turn, the (proxy) stellar phase space density $\propto N_\star \rh^{-3} \langle \sigma^2_\star \rangle^{-3/2}$ rapidly grows. This evolution facilitates the dynamical formation of stellar binaries, as discussed in \citet{EPW2025} for the case of stellar populations that contract in a static smooth dark matter halo. The $N$-body models of the present work build upon the results discussed in \citet{EPW2025}, but allow the exchange of energy between stars and dark matter particles. This pathway for collisional cooling with dark matter particles turns out to be remarkably efficient in facilitating the formation of stellar binaries in dark matter-dominated systems. 

We denote the fraction of binary stars by $f_\mathrm{bin} \equiv N_\mathrm{bin} / N_\star$, defined as the ratio of stars in binary pairs normalized to the total number of stars. We identify stellar binaries as pairwise associations of stars with a negative Keplerian binding energy\footnote{We compute binding energies using the same spline-softened potential as adopted in the $N$-body code (see Sec.~\ref{Sec:NbodyCode} for details).} ${E_\mathrm{bin} < 0}$. We do not double-count stars that are part of multiple pair-wise associations, i.e., $f_\mathrm{bin} \leq 1$. We further require that each binary pair has an orbital period $T_\mathrm{bin}$ shorter than the circular period $T_\mathrm{com}$ of the binaries' center of mass within the dark matter potential. This condition serves to impose a tidal limit, and expressed in terms of densities, it implies that the mean stellar density within the binary semi-major axis exceeds the mean dynamical density within a radius equal to the binaries' center of mass. That is, $2m_\star/a_\mathrm{bin}^3 \ge M_\mathrm{dyn}({<}r_\mathrm{com})/r_\mathrm{com}^3$, where by $M_\mathrm{dyn}({<}r_\mathrm{com})$ we denote the total enclosed stellar and dark matter mass. Binaries that satisfy these criteria are highlighted with red circles in Fig.~\ref{Fig:projected_densities}.

Stellar binaries form and disrupt dynamically in our simulations\footnote{The adopted force softening implies that binaries with separations smaller than $\epsilon_\star=0.02\,\pc$ are not resolved. The corresponding hardness parameter \citep{Heggie1975} is at most of order unity: for the $\Ydynzero=10$ model, setting $a_\mathrm{bin} = \epsilon_\star$ and $m_\star=0.8\,\Msol$ yields $G m_\star/ (2 a_\mathrm{bin} \langle\sigma_\mathrm{los}^2\rangle ) \approx 0.5$.}. The top panel of Fig.~\ref{Fig:Binaries} shows the evolution of the binary fraction $f_\mathrm{bin}$ in the $\Ydynzero=10$ model, for binaries with semi-major axis $a_\mathrm{bin} \leq 0.1\,\pc$. After roughly $2\,\Gyr$ of evolution,  $f_\mathrm{bin}$ rises rapidly, reaching a value of $f_\mathrm{bin}\sim15$ per cent around $t=3\,\Gyr$ and $f_\mathrm{bin}\sim20$ per cent around $t=4\,\Gyr$. Note that these binary fractions are substantially larger than the values reported in \citet{EPW2025}. We attribute this difference to the cooling mechanism enabled by the energy exchange between dark matter and stars, which is accounted for in the present $N$-body work. 

The stellar binaries contribute significantly to the potential energy of the stellar population. This is shown in the bottom panel of Fig.~\ref{Fig:Binaries}, where we plot the time evolution of the potential energy sourced by the stars alone, $W_\mathrm{self} = \sum m_\star \Phi_\star/2$. We compute this potential energy in two different ways: First through direct summation, which defines the total $W^\mathrm{tot}_\mathrm{self}$ (orange curve in the bottom panel of Fig.~\ref{Fig:Binaries}). Second by assuming spherical symmetry, $W^\mathrm{sph}_\mathrm{self} =2 \pi \int \diff r~ r^2 \rho_\star(r) \Phi_\star(r) $, which serves as an estimate of the potential energy of the bulk stellar density distribution (green curve). The difference between these two values serves as an estimate of the potential energy stored in stellar binaries, $W^\mathrm{bin}_\mathrm{self}$ (blue curve). 

As time progresses, the total $W^\mathrm{tot}_\mathrm{self}$ becomes more negative. Initially, this is driven by the contraction of the cluster, and the spherical $W^\mathrm{sph}_\mathrm{self}$ provides virtually all of the total potential energy. However, once binaries form, the evolution of $W^\mathrm{sph}_\mathrm{self}$ flattens off while $W^\mathrm{tot}_\mathrm{self}$ continues to become more negative: stellar binaries contribute increasing amounts of potential energy to the system; in turn, the flattening of $W^\mathrm{sph}_\mathrm{self}$ slows down the contraction of the stellar population\footnote{This result critically depends on adopting a sufficiently-small gravitational softening length for the interactions between stars. We note that when using a softening length of $0.1\,\pc$ between stars, five times larger than the value of $\epsilon_\star=0.02\,\pc$ adopted in this study, the stellar population keeps contracting further, and $W^\mathrm{bin}_\mathrm{self}$ hardly contributes to the total potential energy which remains dominated by $W^\mathrm{sph}_\mathrm{self}$.}. 

\begin{figure}[t]
 \centering
 \includegraphics[width=\columnwidth]{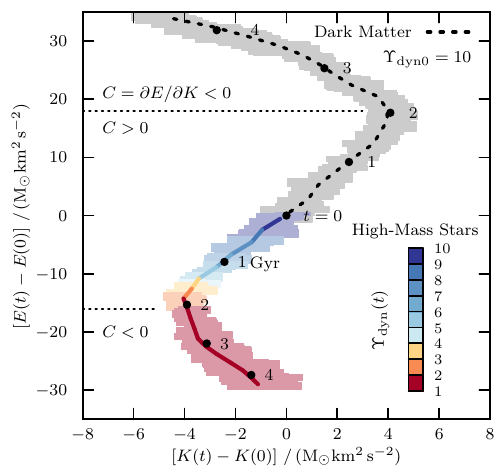}
\caption{Evolution of the total energy $E(t) - E(0)$ and the kinetic energy $K(t) - K(0)$ for dark matter (gray) and high-mass stars (color-coded by $\Ydyn$). As the high-mass stars contract, heat (i.e., kinetic energy) is transferred to the dark matter, and both the stellar and the dark matter systems have a positive heat capacity $\partial E / \partial K >0$. Once the stars become self-gravitating ($\Ydyn \approx 1$), the stellar population heats up as it contracts further: the heat capacity switches sign and becomes negative $\partial E / \partial K <0$.   }
\label{Fig:Thermodynamics}
\end{figure}  

\subsection{Thermodynamics}
\label{Sec:Thermodynamics}
The exchange of heat between stars and dark matter gives rise to a rich phenomenology, where stars initially cool as they contract, and in later stages of the evolution heat up as they contract further. Some intuition into the underlying thermodynamics can be gained by studying the exchange of kinetic and potential energy between stars and dark matter. We compute the total kinetic energy of the stars as $ K_\star = \sum m_\star v_\star^2/2 $, where the sum is performed over all stars. Equivalently, for the dark matter, $ K_\mathrm{DM} = \sum m_\mathrm{DM} v_\mathrm{DM}^2 / 2$.
We then define the potential energy of the stars in the combined potential as $W_\star \equiv \sum m_\star (\Phi_\star + \Phi_\mathrm{DM})/2$ and equivalently $W_\mathrm{DM} \equiv \sum m_\mathrm{DM} (\Phi_\mathrm{DM} + \Phi_\star)/2$. With this definition, $W_\star$ and $W_\mathrm{DM}$ symmetrically share the star--dark matter cross terms of the binding energy.
The total conserved energy of the combined system of stars and dark matter reads $E_\mathrm{tot} = \text{const} = E_\star + E_\mathrm{DM}$ where $E_\star \equiv K_\star + W_\star$ and $E_\mathrm{DM} =  K_\mathrm{DM} + W_\mathrm{DM}$. The heat capacity of the stars, computed in the combined gravitational potential\footnote{For the case of an isolated and virialized system of stars in absence of dark matter ($\Phi_\mathrm{DM}=0$, i.e. $W_\star \equiv W_\mathrm{self}$), the definitions above recover the classical $C_\star=1+W_\star/K_\star=-1$, as for a self-gravitating system in virial equilibrium, potential energy $W_\star$ and kinetic energy $K_\star$ are related through $W_\star = - 2 K_\star$. } of stars and dark matter, can then be written as $C_\star \equiv \partial E_\star / \partial K_\star$, and equivalently, $C_\mathrm{DM} \equiv \partial E_\mathrm{DM} / \partial K_\mathrm{DM}$.
For the sake of numerical efficiency, for the dark matter, we compute the potential energy term under the assumption of spherical symmetry, whereas for the stars, we rely on direct summation. 

Figure \ref{Fig:Thermodynamics} shows the change in kinetic energy $\Delta K = K(t) - K(0)$ as a function of the change in total energy $\Delta E = E(t) - E(0)$ for high-mass stars, color-coded by $\Ydyn(t)$, and for dark matter, in gray, for the simulation with $\Ydynzero=10$. As the high-mass stars become more tightly bound within the combined potential ($E_\star$ decreases monotonically), the dark matter becomes less bound ($E_\mathrm{DM}$ increases). In the early phase of the evolution, stars are gravitationally sub-dominant. The heat capacity of the high-mass stars is positive, $C_\star > 0$. Consequently, as the high-mass stars contract and become more gravitationally bound, they cool down. Once the high-mass stars become gravitationally dominant within their own luminous radii, the heat capacity changes sign and becomes negative, $C_\star < 0$. In turn, as the high-mass stars continue to contract and become more tightly bound, they heat up. In the simulation studied here, the population of low-mass stars hardly contributes to this exchange of energy. As the total energy of the system is conserved, the evolution of the dark matter particles in the $\{\Delta E,\Delta K\}$ plane approximately mirrors that of the high-mass stars.

\section{Discussion}
\label{Sec:Discussion}
\textit{Summary.} We study the collisional exchange of energy between stars and dark matter in tidally limited ultra-faint galaxies. Our $N$-body simulations adopt dark matter particles with a mass of $m_\mathrm{DM}=10^{-3}\,\Msol$ and a two-component stellar population consisting of high- and low-mass stars ($m_\star=0.8$ and $0.2\,\Msol$, respectively). We show that, as dynamical friction with dark matter causes high-mass stars to sink to the galaxy's center, the initial central dark matter cusp is transformed into a constant-density core. During this process, both stars and dark matter have positive heat capacity: the stellar component cools as it contracts. The dynamical-to-stellar mass ratio within the half-light radius of the high-mass stellar population decreases monotonically, transforming a dark matter-dominated stellar system into a dense, self-gravitating star cluster, surrounded by a dark matter halo. Once stellar self-gravity dominates over dark matter, the heat capacity switches sign and becomes negative: the star cluster then heats up as it contracts further. The population of low-mass stars contracts much more slowly than the population of high-mass stars, resulting in stellar mass segregation driven by dynamical friction. The mere segregation of stellar masses therefore cannot serve as a litmus test for the absence of dark matter. The contraction of the population of high-mass stars is eventually decelerated due to the decreasing central dark matter densities and the formation of stellar binaries. 

\textit{Observational consequences.} 
Our $N$-body models show that small stellar systems with sizes of $\mathcal{O}(10\,\pc)$, if embedded in a cuspy tidally-stripped dark matter subhalo of low central velocity dispersion $\mathcal{O}( 1\,\kms )$, contract due to dynamical friction acting on individual stars. Our models therefore predict that objects like Delve~1, if indeed embedded in a dark matter subhalo, should have formed self-gravitating star clusters consisting of their higher-mass stars at their centers. For systems like Delve~1, the most massive stellar remnants are argued to be white dwarfs. Echoing the results obtained for dark matter-free star clusters by \citet{Devlin2025}, our models suggest that, even in the presence of large amounts of dark matter, systems like Delve~1 likely host central clusters of white dwarfs. We expect the same systematics to apply to the central regions of ultra-faint galaxies with low velocity dispersion, such as Bo\"otes~2 ($\sigma_\mathrm{los} \approx (1.9 \pm 0.8)\,\kms $, see \citealt{Geha2026}), Tucana~3 ($\sigma_\mathrm{los} \lesssim 1.5\,\kms$, see \citealt{Simon2017}), as well as to candidate systems for micro galaxies with similar kinematics, such as Unions~1 ($\sigma_\mathrm{los} \lesssim 2.3\,\kms$, see \citealt{Smith2024,Cerny2026Unions}).

Chemically, the self-gravitating central cluster of massive stars would be expected to mirror the abundances seen in field stars of ultra-faint dwarf galaxies. Signatures in the elemental abundances typical of star formation in high-density systems \citep{Venn2004, Ji2019_Gru1} would therefore be absent in clusters that formed through the contraction of a field population in the presence of dark matter. Such stellar clusters would plausibly have the metallicity dispersion typical of a dwarf galaxy, while having sizes, densities, and kinematics akin to those of globular clusters.

Our controlled simulations (based on an idealized two-component stellar mass function) suggest that ${\sim}$20 per cent of the high-mass stars ($m_\star =0.8\,\Msol$) in the central cluster are part of dynamically formed binary pairs with semi-major axes $a_\mathrm{bin} \leq 0.1\,\pc$ (or, of higher-order associations). The statistics of (wide) binaries have been proposed as a sensitive probe for dark matter substructures \citep{PenarrubiaLudow2016, Shariat2025}, and our simulation results highlight the necessity for detailed modelling of the dynamical interaction between stars and dark matter to accurately estimate the abundances of wide binaries in ultra-faint galaxies. While the kinematic detection of (wide) binaries is highly challenging because of the low radial velocities involved, a population of wide binaries may be detected by its imprints on the stellar two-point correlation function \citep{LonghitanoBinggeli2010, Kervick2022, Safarzadeh2022, Shariat2025}.

The dark matter halos studied here initially follow centrally-divergent (cuspy) density profiles. After $2\,\Gyr$ of evolution, however, we find that they develop constant-density cores, with core sizes (see Eq.~\ref{Eq:Core}) of $r_\mathrm{c}/r_\mathrm{s} \approx 0.2$, $0.4$ and $1.0$ for the models with $\Ydynzero = 30$, $10$ and $3$, respectively. This cusp-to-core transformation driven by dynamical friction opens a new pathway for the formation of density cores in systems where the effects of baryons on the dark matter distribution were traditionally expected to be negligible \citep{Penarrubia2012, DiCintio2014, Onorbe2015}. This, in turn, will decrease the astrophysical $J$-factor $\propto \langle \rho^2_\mathrm{DM} \rangle$ of a possible dark matter self-annihilation signal \citep[see, e.g.,][]{DiemandKuhlenMadau2007, Walker2011_Jfactor, Moline2017, Crnogorcevic2024, ENSM24}.
 
\textit{Caveats.} Our study makes various simplifying assumptions, which we discuss below.
(1)~The dark matter halo surrounding the ultra-faint galaxy does not contain any substructure (subhalos, sub-subhalos, etc.). Analogous to the heating of stellar streams \citep{Ibata2002heating, Johnston2002}, collisional heating by dense dark matter substructures would inject energy into the stellar population \citep{Penarrubia2025}, an effect that we neglect in the current study. Dark matter mini-halos are argued to be strongly stripped or even fully disrupted by direct collisions with stars \citep{Delos2019, Fachinetti2022}, hence, we consider the detailed substructure abundance within ultra-faint galaxies an open question that requires further study. 
(2)~Dark matter cusps have been shown to re-form by means of minor mergers \citep{LaportePenarrubia2015}, which we do not account for. However, for the tidally limited systems studied here, accretion of dark matter onto the ultra-faint galaxy's subhalo likely ceased with its own accretion onto the Milky Way.
(3)~We adopt an idealized two-component stellar population consisting of high- and low-mass stars ($m_\star=0.8$ and $0.2\,\Msol$, respectively), without stellar evolution. This choice is made to facilitate the interpretation of the simulation results, but inevitably reduces the realism of our models. A spectrum of stellar masses would allow for energy exchange between stars of different mass, resulting in stellar mass stratification. Further, a more realistic mass function plausibly affects the detailed number of dynamically formed stellar binaries.
(4)~Our initial conditions are spherically symmetric and assume isotropic kinematics, both for the stellar populations and for the dark matter. Our models do not contain any primordial stellar binaries, which -- in addition to the population of dynamically formed binaries -- may serve as sinks for potential energy and contribute to the total velocity dispersion. 
(5)~Crucially, we adopt force softening of $\epsilon_\star = 0.02\,\pc$ for the interactions between stars. We therefore do not resolve the formation of close binaries within the stellar cluster. Our setup is not designed to accurately follow the secular evolution of the star cluster once it becomes self-gravitating. We therefore omit a detailed description of the star cluster structure and kinematics, and defer this analysis to follow-up work. 

\textit{Outlook.}
Our results show that collisional processes play a key role in shaping the evolution of the faintest, smallest galaxies. A detailed understanding of the collisional exchange of energy between stars and dark matter in ultra-faint galaxies will be crucial for determining the formation rates of wide binaries and the merger rates of heavy stellar remnants. We aim to address these questions in future work.

\section*{Acknowledgments}
RE would like to thank Neal Dalal and Ling Tan for stimulating discussions. RE and MW acknowledge support from the National Science Foundation (NSF) grant AST-2206046. Support for program JWST-AR-02352.001-A was provided by NASA through a grant from the Space Telescope Science Institute, which is operated by the Association of Universities for Research in Astronomy, Inc., under NASA contract NAS 5-03127. This material is based upon work supported by the National Aeronautics and Space Administration under Grant/Agreement No. 80NSSC24K0084 as part of the Roman Large Wide Field Science program funded through ROSES call NNH22ZDA001N-ROMAN. NE is a FRIA grantee of the Fonds de la Recherche Scientifique-FNRS (ULB-TH/26-06) and a member of BLU-ULB.

\section*{References}
\bibliographystyle{aasjournal-hyperref}
\bibliography{friction}

\label{lastpage}
\end{document}